\documentclass[aps,prl,groupedaddress,superscriptaddress,twocolumn,10pt,longbibliography]{revtex4-2}
\pdfoutput=1 
\usepackage[utf8]{inputenc}
\usepackage{hyperref}
\usepackage{graphicx}
\usepackage{setspace}
\usepackage{orcidlink}
\usepackage{footnote}
\usepackage{soul}
\newcommand{\corauthor}[2]{
    \author{#1}
    \email{#2}
}

\usepackage{siunitx} 

\usepackage[caption=false]{subfig}
\usepackage{tabularray}
\usepackage{amssymb}
\usepackage{physics}
\usepackage{float}
\DeclareSIUnit\bohr{\text{\ensuremath{a_\textup{0}}}}

\begin{document}

\title{High precision spectroscopy of trilobite Rydberg molecules}

\author{Markus Exner\orcidlink{0009-0005-8290-7371}}
\thanks{These authors contributed equally to this work.}
\affiliation{Department of Physics and Research Center OPTIMAS, Rheinland-Pfälzische Technische Universität Kaiserslautern-Landau, 67663 Kaiserslautern, Germany}

\author{Rohan Srikumar\orcidlink{0000-0003-0303-1331}}
\thanks{These authors contributed equally to this work.}
\affiliation{Zentrum für Optische Quantentechnologien, Universität Hamburg, Luruper Chaussee 149, 22761 Hamburg, Germany}

\author{Richard Blättner\orcidlink{0009-0004-4667-821X}}
\affiliation{Department of Physics and Research Center OPTIMAS, Rheinland-Pfälzische Technische Universität Kaiserslautern-Landau, 67663 Kaiserslautern, Germany}

\author{Matthew T. Eiles\orcidlink{0000-0002-0569-7551}}
\affiliation{Max Planck Institute for the Physics of Complex Systems,  Nöthnitzer Str. 38, 01187 Dresden, Germany}

\author{Peter Schmelcher\orcidlink{0000-0002-2637-0937}}
\affiliation{Zentrum für Optische Quantentechnologien, Universität Hamburg, Luruper Chaussee 149, 22761 Hamburg, Germany}
\affiliation{The Hamburg Centre for Ultrafast Imaging, Universität Hamburg, Luruper Chaussee 149, 22761 Hamburg, Germany}

\corauthor{Herwig Ott\orcidlink{0000-0002-3155-2719}}{Corresponding author: ott@physik.uni-kl.de}
\affiliation{Department of Physics and Research Center OPTIMAS, Rheinland-Pfälzische Technische Universität Kaiserslautern-Landau, 67663 Kaiserslautern, Germany}

\date{\today}

\begin{abstract}
We perform three-photon photoassociation to obtain high resolution spectra of $^{87}$Rb trilobite dimers for the principal quantum numbers $n = 22,24,25,26$, and 27. The large binding energy of the molecules in combination with a relative spectroscopic resolution of $10^{-4}$ provides a rigorous benchmark for existing theoretical models. A recently developed Green’s function framework, which circumvents the convergence issues that afflicted previous studies, is employed to theoretically reproduce the vibrational spectrum of the molecule with high accuracy. The relatively large molecular binding energy are primarily determined by the low energy $S$-wave electron-atom scattering length, thereby allowing us to extract the $^3S_1$ scattering phase shift with unprecedented accuracy, at low energy regimes inaccessible to free electrons.



\end{abstract}

\maketitle


\paragraph{Introduction:} The ability to introduce Rydberg excitations in ultracold atomic gases has set in motion a new branch of molecular physics, wherein atoms are bound over large distances via unconventional binding mechanisms \cite{Greene2000,Rosario_2015,Mayle_2012,Hollerith_2023,Raithel_2021,Dan_2023,Zuber_2022}.
Trilobite molecules are a peculiar category of such ultralong-range Rydberg molecules (ULRM) \cite{Greene2000,Eiles_2019,Fey_2020_review,Dunning_2024_review,Hamilton_2002}, bound by a ground state atom scattering off a Rydberg electron with high angular momentum $\ell$. These homonuclear dimers, which exhibit permanent electric dipole moments in the kilo-Debye range \cite{Li2011,Chibisov_2002,Booth2015,Althoen2023} and have bond-lengths of the order of micrometers, promise dynamical effects and field control not typical of conventional molecules \cite{engel2023situ,trilobite_wave_packet,srikumar2024}. Although the first experimental evidence for the trilobite dimer was obtained by mixing the trilobite wavefunction to low-$\ell$ Rydberg states \cite{Booth2015,PhysRevLett.118.223001}, it was only very recently that direct photoassociation of the molecule, accessing the trilobite potential energy curves (PECs) was made possible \cite{Althoen2023}.
For more than a decade, numerous spectroscopic studies have been performed on ULRM \cite{Bendkowsky2009,PhysRevLett.123.073003,Spinflip,butterfly,PhysRevLett.114.133201,Deiss_2020,PhysRevA.102.062819,PhysRevLett.126.013001,PhysRevA.99.033407} providing detailed insights into the spin-structure, potential energy landscape, and rovibronic spectra of these dimers. 
Since the low-energy scattering of the Rydberg electron forms the binding mechanism of the molecule, the depth of the potential wells and the vibrational binding energies are directly dependent on the size of the scattering partial wave $L$.
In turn, high resolution spectroscopic data can provide precise values of the spin-dependent scattering lengths, ultimately testing the validity of the underlying interaction mechanism. 
Consequently, the ULRM has been used as a laboratory to extract electron-atom scattering lengths and resonance positions at low energy regimes inaccessible with free electrons \cite{PhysRevLett.123.073003}.

In this work, we combine the newly formulated Green's function method \cite{green_function} with the advantages of pure trilobite photoassociation spectroscopy. Due to the large molecular binding energies and the high precision of spectroscopic measurement, we determine the potential energy depth with unprecedented accuracy, and extract the energy-dependent $e^- -$ Rb, $^3S_1$ scattering length. At the same time,  we find indications that the predictive power of the underlying molecular Hamiltonian is being challenged, and further insight in the direction of non-adiabatic physics \cite{VibronicRb,VibronicNa,KatosMatt} and/or higher order scattering terms is necessary before we can unravel agreement between experiment and theory data within the resolution achieved.

\paragraph{Interactions and methodology:} The Born-Oppenheimer PECs $U_i(R)$, and the associated electronic states $\phi_{i}(\vec r; R)$ are obtained by solving $H\phi_{i}(\vec r; R) = U_i(R)\phi_{i}(\vec r; R)$, where the electronic Hamiltonian $H$ incorporates the fine-structure of the Rydberg electron, hyperfine-structure of the ground-state atom (with a total spin $F$=1,2 for $^{87}$Rb), and the six different electron-atom scattering channels ($^1S_0,^3S_1,^1P_1,^3P_{0,1,2}$). Until recently, the standard method for determining the PECs was the diagonalization of $H$ in a truncated basis \cite{Eiles_2017}.
Here, the scattering interaction between the Rydberg electron and perturber is characterized by a Fermi-type pseudopotential in $S$-wave \cite{Fermi1934} and $P$-wave channels \cite{Omont1977}.
While providing a valid semi-quantitative description of the system, the Dirac-delta form of the pseudopotential introduces unphysical basis size dependence of the PECs obtained utilizing diagonalization schemes \cite{Fey_2015}. Furthermore, the approach suffers from diverging scattering volumes in the $^3P_J$ scattering channels introduced by negative ion resonances.
For this reason, the scattering phase shift extracted by comparing experimental and theoretical binding energies obtained utilizing this method needs to be handled with care. Alternative approaches have been developed, but as these do not include all spin interactions, they are not sufficiently accurate to meet high-precision spectroscopy \cite{PhysRevA.102.062802,PhysRevA.102.033315}.
However, the recently developed non-perturbative Green's function formulation (see \cite{green_function} for  detailed explanation) accounts for all spin interactions, and circumvents the aforementioned convergence issues.

\begin{figure}
\centering
\includegraphics[scale=1]{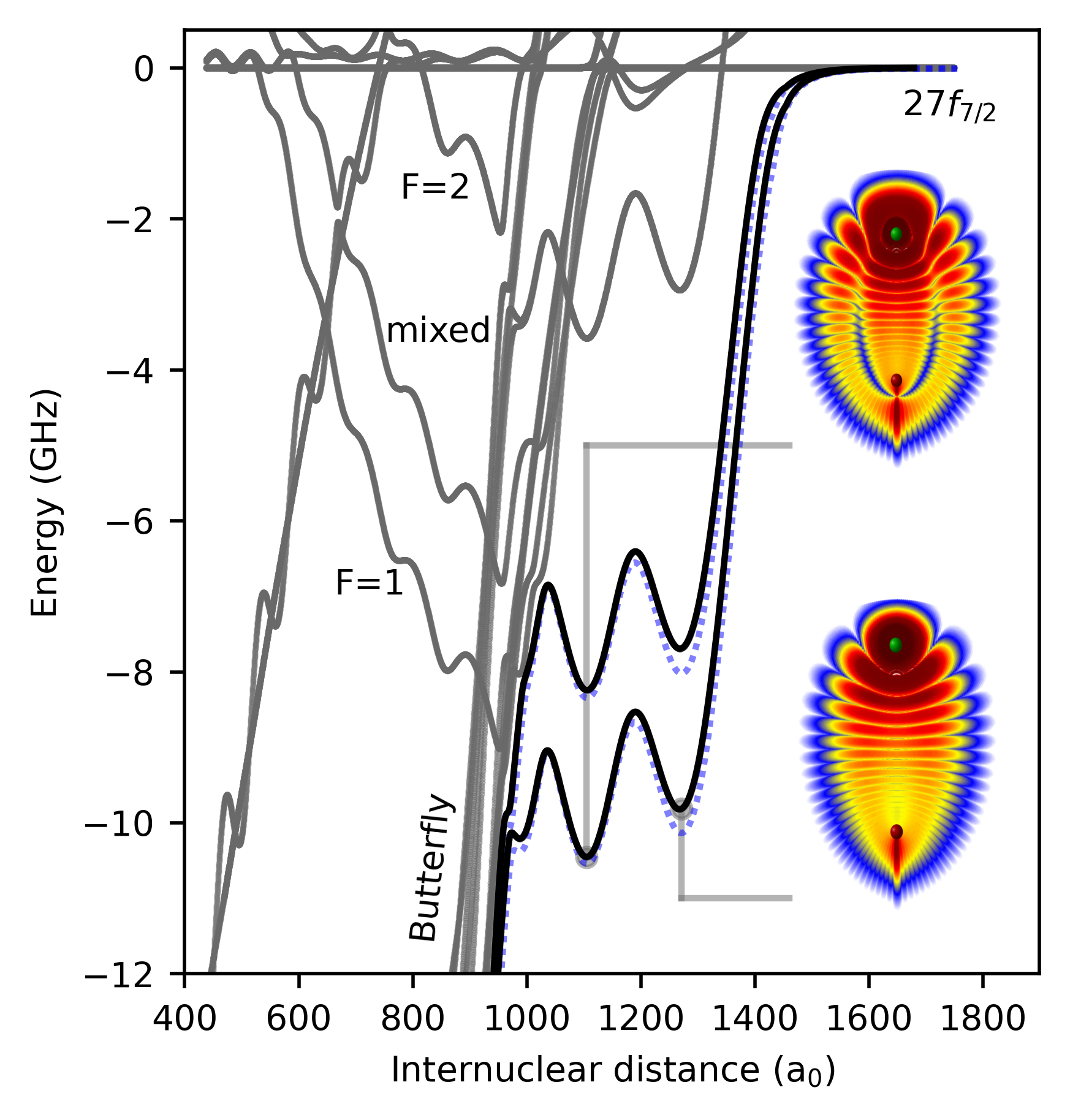}

\caption{Rydberg $n$=$27$, $M$=$1/2$ molecular Born-Oppenheimer potential curves obtained using the Green's function treatment (grey). The black curves highlight the trilobite potentials we experimentally probe, and for comparison the blue curves are potentials calculated by numerical diagonalization (using $n \in [25,29]$ basis). On the right are two density plots of the Coulomb Green's function which represents the Rydberg electron distribution. Due to the $S$-wave scattering and the admixture of high angular momentum states, the probability of the electron staying at the position of the ground state atom 
is maximised.}

\label{fig:trilobite}
\end{figure}

Figure \ref{fig:trilobite} shows the Born-Oppenheimer potential energy landscape of the ULRM obtained using the Green's function calculation for principal quantum number $n$=27, with the projection of total angular momentum along the internuclear axis taken to be $M$=$1/2$.
The zero frequency is the asymptotic pair state Rb($27 f_{7/2}$) + Rb(5s,$F=1$), from which the two potential curves emerge (black), corresponding to the pure triplet scattering channel (lower-PEC), and a singlet-triplet mixed channel (upper-PEC).  For illustration the same two potentials were calculated with the diagonalisation method (blue dotted), which leads to \SI{10}{\percent} deviation in the binding energy.
At \SI{1050}{\bohr} the trilobite levels interact with the plummeting ``butterfly" potential curves resulting from the shape resonance in the $P$-wave scattering, of which there are five of predominantly $^3P_2$ configuration, three of $^3P_1$, and a single $^3P_0$. In this work we experimentally access the vibrational bound states in the highlighted trilobite curves (hereby referred to as triplet trilobite and mixed trilobite PEC)  for $n = 22,24,25,26,$ and $27$
\newline

\paragraph{Experimental setup:}We prepare the atomic sample starting from a magneto-optical trap (MOT) of $^{87}$Rb atoms, which are subsequently transferred into a crossed dipole trap operating at $\lambda = \SI{1064}{nm}$. The spherical atomic ensemble has a temperature of \SI{40}{\micro K}, a diameter of 40\,$\mu m$ and a peak density of \SI{4 e13}{cm^{-3}}. The atoms are prepared in the $F = 1$ hyperfine ground state.
Employing a three-photon excitation scheme (5$s_{1/2}\to $5$p_{3/2} \to $5$d_{5/2} \to$$nf_{7/2}$) at wavelengths of \SI{780}{nm}, \SI{776}{nm}, and \SIrange{1276}{1288}{nm}, we can photoassociate pure trilobite Rydberg molecules utilizing its $f$-state admixture. For precise frequency control, all three excitation lasers are frequency stabilized to an ultra-low expansion cavity using a Pound-Drevel-Hall locking technique.
Immediately following the excitation, the Rydberg atoms are ionized using a CO$_2$ laser. The resulting ions are guided by an electric field in a Wiley-McLaren configuration onto a time- and position-resolved multichannel plate detector, enabling the direct measurement of the momentum distribution prior to ionization. 
Hence, we are able to distinguish the ions that have zero momentum, and consequently detect long-lived stable molecular states \cite{Althoen2023}. The full momentum signal however, contains the signatures of both long-lived and short-lived molecular resonances.
The generated ions are counted for each detuning step (\SI{2}{MHz}) and averaged over typically 10 experimental runs. To estimate the maximum error for the frequency, we use the accuracy of the free spectral range of \SI{100}{kHz}, resulting in an error for the lowest trilobite state of $\frac{\SI{21}{GHz}}{\SI{1.4978}{GHz}} \cdot \SI{100}{kHz} \approx \SI{1.4}{MHz}$. On the other hand, we use the frequency step size of \SI{2}{MHz} as the readout error. This results in a frequency uncertainty of \SI{3.4}{MHz}, which corresponds to a deviation smaller than \SI{0.01}{\percent} at a binding energy of \SI{21}{GHz} and thus illustrates the excellent accuracy of the experiment.

\begin{figure}
\centering
\includegraphics[width=\linewidth]{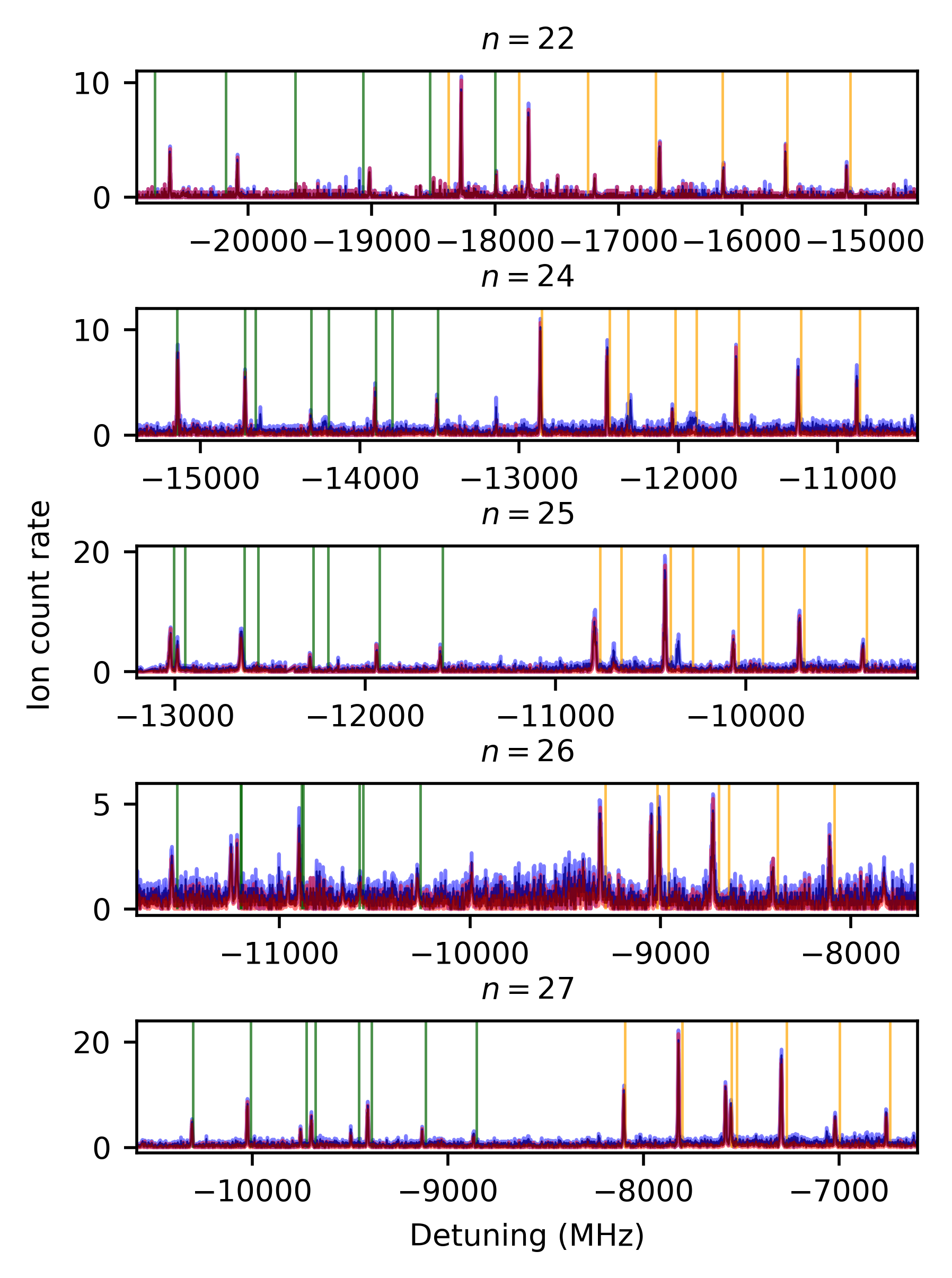}
\caption{The zero momentum (red) and full momentum (blue) ion signals as a function of the frequency detuning from the $nf_{7/2}$ atomic resonance, for all measured $n$-values. The calculated eigenenergies are shown as vertical lines for the triplet $F=1$ (green) and the mixed (orange) potential  }
\label{fig_spectrum}
\end{figure}

\begin{figure}
\centering
\includegraphics[scale=0.85]{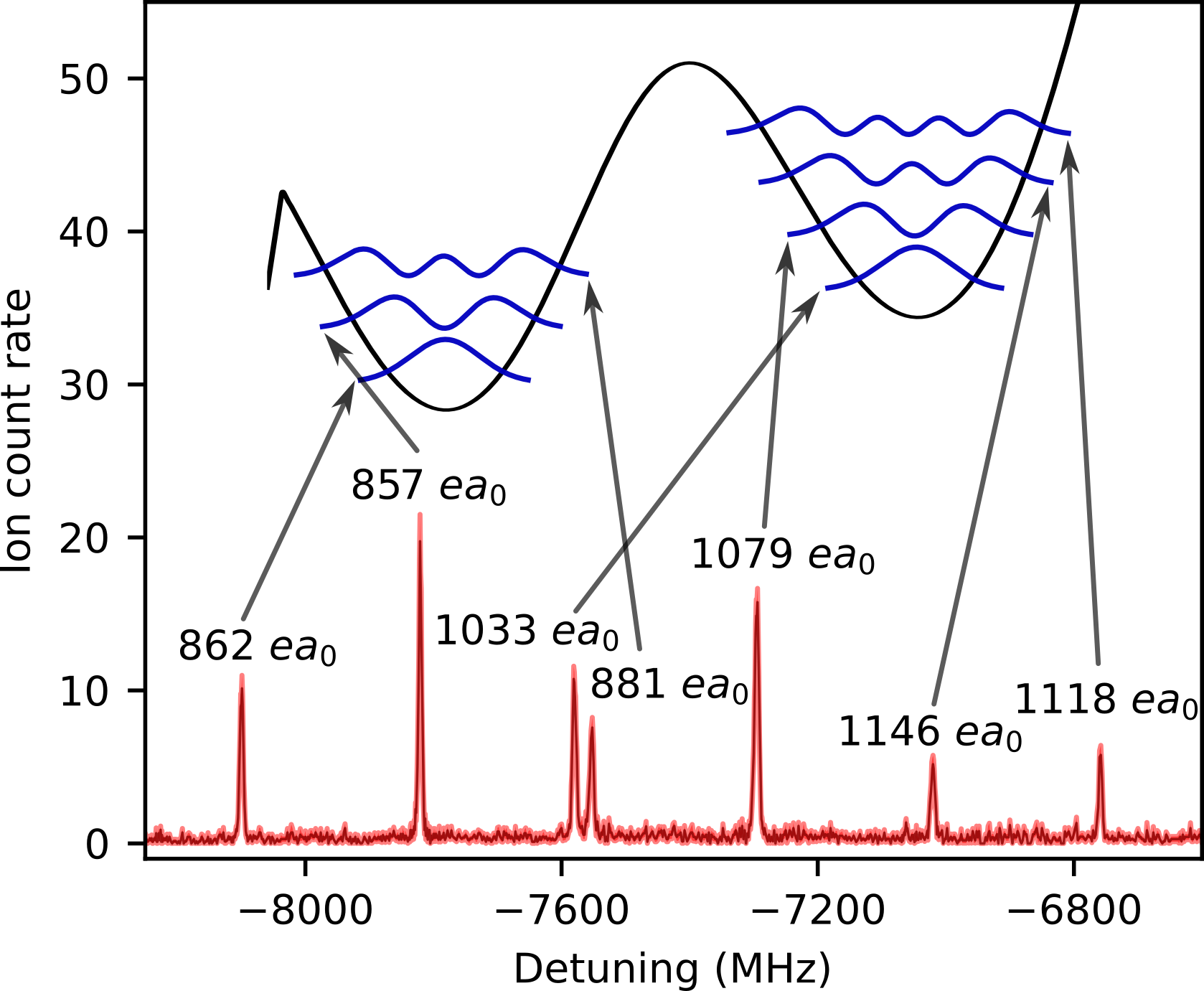}
\caption{\footnotesize{The trilobite spectrum (red) for the energy range of the mixed potential for a principal quantum number of n=27 with the potential energy curve (black) above including vibrational probability densities (blue, not to scale). The observed dipole moments of each measured peak is shown and assigned to the corresponding vibrational state.}}
\label{fig:exp_27F}
\end{figure}

\paragraph{Spectral analysis:}  Figure \ref{fig_spectrum} shows the observed spectra for all measured $n$, which are several GHz red-detuned from the $nf_{7/2}$ atomic resonance.
Two different vibrational series can be identified, which still overlap at the principal quantum number of 22 but are clearly separated at $n=25$. These originate from the mixed and the triplet $F=1$ PECs, respectively.
The vertical lines show the calculated binding energies for the triplet $F=1$ (green) and the mixed (orange) trilobite states.
The number of vibrational states in each potential is limited by the position of the crossing with the butterfly PECs, as tunneling towards shorter internuclear distances occurs, leading to state-changing collisions or associative ionization \cite{Althoen2023,State_changing}. 
For $n < 24$ there is a single outer well beyond the $P$-wave crossing, whereas for $n$=24 and above, there is a second potential well. This explains the presence of double peaks in the spectrum  for $n$ =24-27 that are absent in the $n$=22 spectra (See End Matter section for more details).
The presence of the second potential well renders the assignment of the vibrational states challenging. Here, knowledge of the permanent dipole moment helps in the identification of states, as the dipole moment of the vibrational state increases with bond-length. To determine the dipole moment, a weak electric field is applied and the broadening of the state is measured \cite{Althoen2023}. The measured dipole moments for different vibrational states of the $n$=27 mixed trilobite potential is shown in Fig. \ref{fig:exp_27F}. Their magnitudes accumulate around two different values, which clearly allows the assignment of the vibrational states to the two calculated potential wells with minima at \SI{1100}{\bohr} and \SI{1275}{\bohr}.
Note that the relative motional state of two ground-state atoms in our setup varies slowly over the extent of the molecular vibrational states, implying minimal Franck-Condon overlap for odd molecular vibrational states. However, the molecular spectra shows that both odd and even vibrational states have comparable signal strength.
This phenomenon is explained by considering the oscillations of the $nf_{7/2}$ admixture in the trilobite state (responsible for the dipole transition from 5$d_{5/2}$), that violates the Franck-Condon approximation that electronic transitions are independent of the nuclear coordinates and facilitates the photoassociation of odd vibrational states.

 \begin{figure}
\centering
\includegraphics[width=\linewidth]{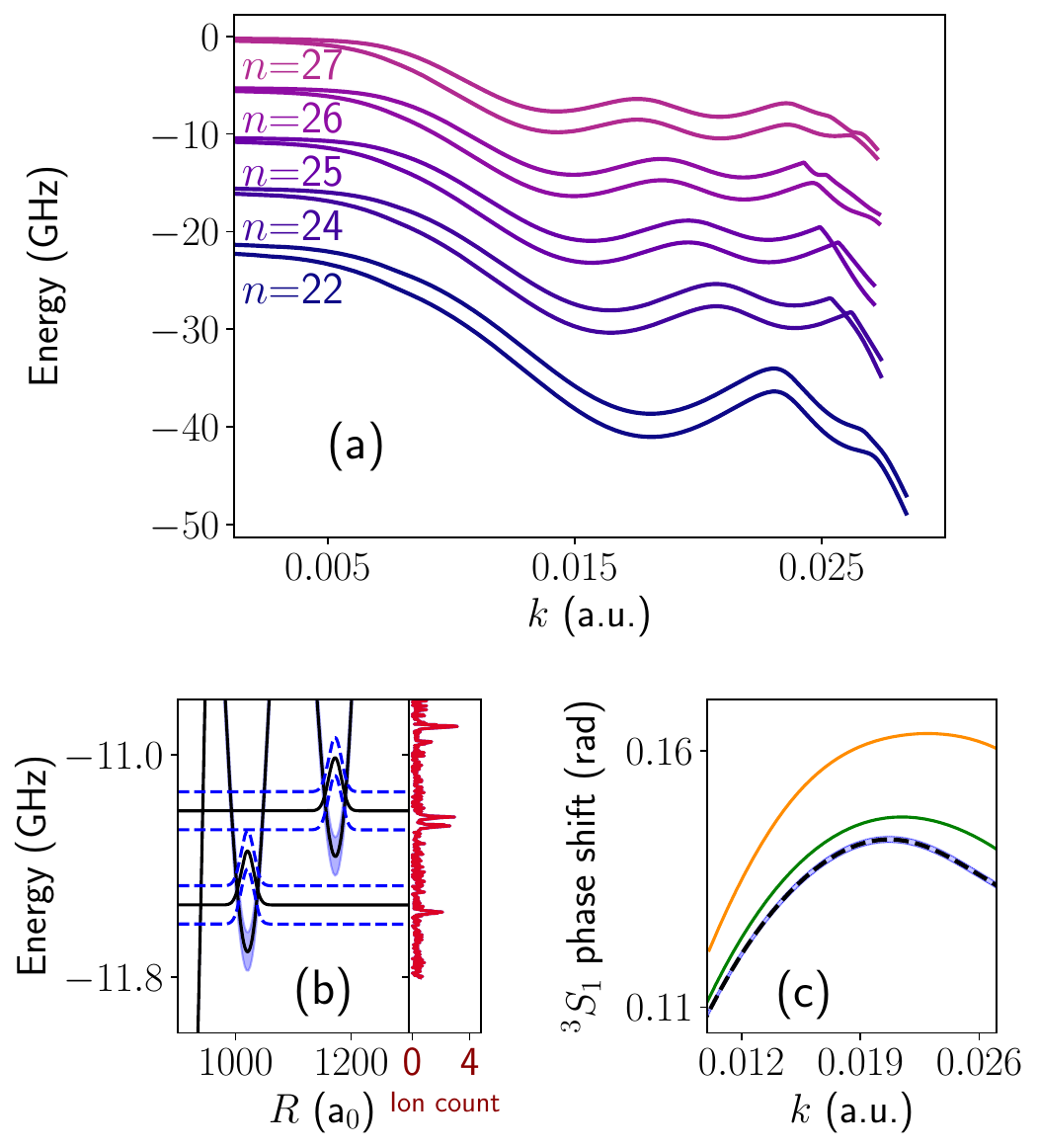}

\caption{Potential energy structure and extracted $S$-wave phase shifts. (a) Illustrative sketch of the probed trilobite PECs as a function of $k$, limited between the energy zeros ($k$=0) and the $P$-wave crossings ($k\sim$0.025) for each PEC. (b) The PECs (solid blue/black lines) and the ground vibrational states (dashed) of the molecule, for n=26, in comparison with the ion-count spectra (red). (c) The dashed black curve shows the $^3S_1$ phase shifts fitted anew, with the shaded blue region in the background portraying the uncertainty in the fitting of phase shifts (see text), leading to corresponding deviations in binding energy as shown in (b). The green curve is the phase shift taken from the previous fit (from Ref.~\cite{PhysRevLett.123.073003}), and the orange curve is the ab-initio calculation of the phase-shift (from Ref.~\cite{Khuskivadze2002}).
}
\label{fig:phase}
\end{figure}

\paragraph{Phase shift extraction:} We now discuss the extraction of the low-energy $S$-wave scattering phase shift from the high resolution data experimentally obtained.
By solving the un-physical basis-dependencies that afflicted other calculations, we are able modify and correct previous phase shifts in such a way that it reproduces the observed resonances more precisely, without dependence on any external model parameters.
For this, we added a ninth order polynomial 
to the phase shifts fitted in \cite{PhysRevLett.123.073003} and treated the coefficients as fit parameters. These were varied until the total difference between the theoretical and experimental bound state spectra from all $n$ were simultaneously minimized, while avoiding any  unphysical oscillatory behavior.
Figure \ref{fig:phase} (a) illustrates the trilobite potential energy curves as a function of the semiclassical electron momentum $k$. Note that $k(R)=\sqrt{ 2  U_n(R) + 1/R}$ is calculated separately for each particular PEC $U_n(R)$, and the curves are vertically offset for an easier visual comparison.
The overlapping $k$-ranges for multiple $n$-values provide several data points (vibrational states) in each $k$-window whose energetic positions are sensitive to the changes in phase shift.
By minimizing the error between the computed and observed binding energies for the vibrational ground states localized in the outer-well, for all values of $n$ simultaneously, we prevent possible overfits in our phase shift calculations (See End Matter section for more details).
Figure \ref{fig:phase} (b) shows the trilobite potential energy curves for $n$=26, calculated utilizing the phase shifts shown in Fig. \ref{fig:phase} (c). 
The newly fit $S$-wave scattering phase shift (dashed black line, Fig. \ref{fig:phase} (c)) differs considerably from those previously determined by Ref.~\cite{PhysRevLett.123.073003} or calculated ab-initio by Ref.~\cite{Bahrim2000,Bahrim2001,Khuskivadze2002}.
The shaded blue region in Fig. \ref{fig:phase} (b) and (c) is used to illustrate the sensitivity of the molecular binding energy on the $^3S_1$ scattering length. A $0.8 \%$ change in the $S$-wave phase shift in the relevant $k$-range (Fig. \ref{fig:phase} (c)) induces an error of $\sim 138$ MHz in the binding energy of the $n=26$ vibrational ground state (dashed lines Fig. \ref{fig:phase} (b), nearly two orders of magnitude larger than the experimental resolution.
The uncertainty of $0.7 \%$ in the $^3S_1$ phase-shift is barely visible in comparison to the large deviation from the previously fitted or calculated phase shifts. 
Apart from illustrating the sensitivity of our calculations, the shaded region also shows the range of $S$-wave phase shifts within which the theoretical and experimental data can be selectively matched perfectly for each $n$-value (case of overfitting).
The inability to match observed and calculated binding energies simultaneously for all values of $n$, within experimental error bars, implies that there might be limitations in the description of the molecular interactions in the Hamiltonian. This necessitates a quantitative analysis of the error between the observed spectra and the numerical spectra calculated with the optimized phase shifts.
\newline

\begin{figure}
\centering
\includegraphics[width=1.0\linewidth]{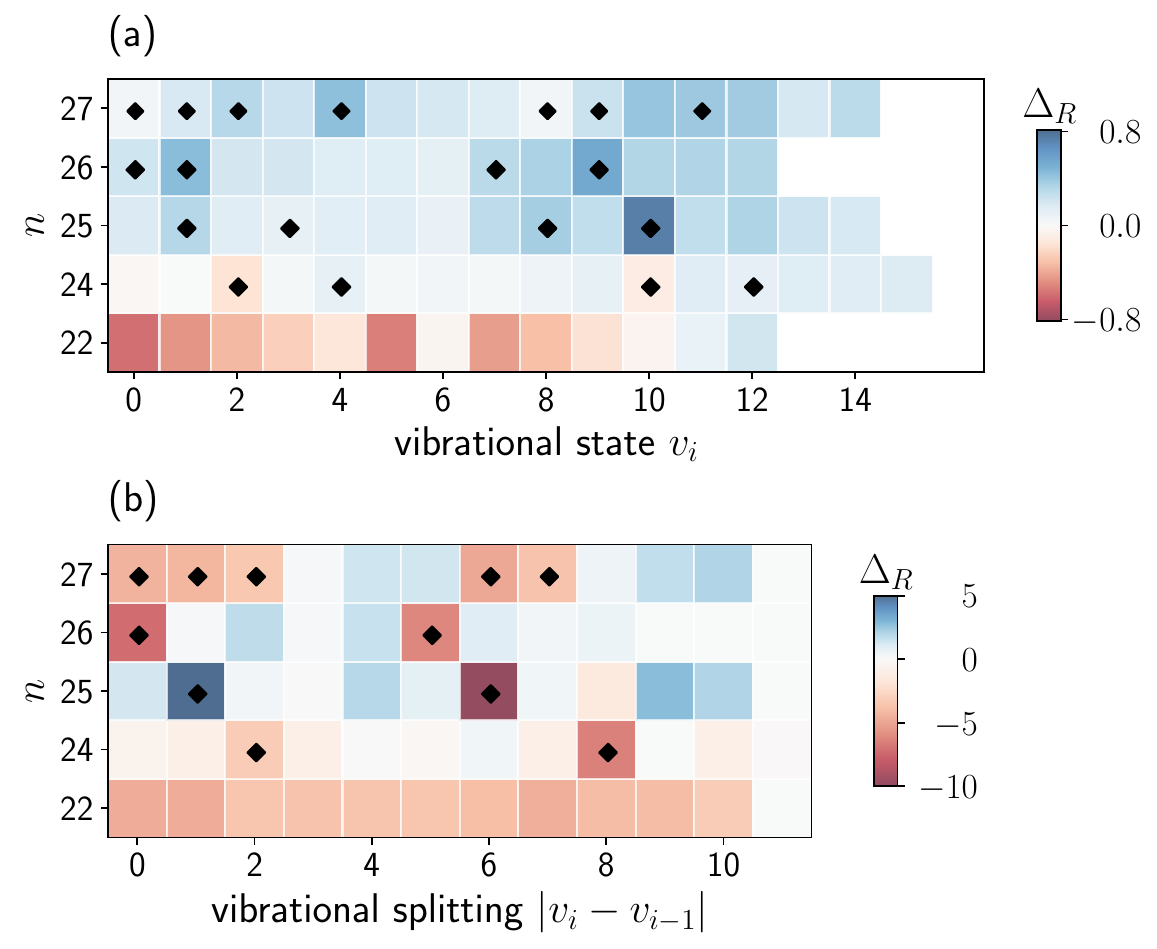}

\caption{The percentage relative difference of the experimental results with respect to the numerical calculations given by $\Delta_R = 100\% \cdot (\epsilon_e - \epsilon_t) / \epsilon_t $  where $\epsilon_e$ and $\epsilon_t$ represent the experimental and theoretical data respectively for (a), the calculated vibrational binding energies and (b), the splitting between vibrational states localised in separate wells. The states localised in the inner-well are marked `$\blacklozenge$'. } 
\label{fig:error}
\end{figure}

Figure \ref{fig:error} (a) portrays the relative error of the binding energies of all the measured vibrational states with respect to their theoretical counterparts.
The vibrational states localized in the outer-wells of the $n\geq24$ trilobite PECs are uniquely $S$-wave dominant and show an excellent maximum relative error $\leq 0.4 \%$.
Although the inner-well states (marked $\blacklozenge$) are also $S$-wave dominant, the exact positions of the  $^3P_{0,1,2}$ crossings finely affects the number of bound states observed. However, we see that their binding energies are also largely dependent on the $^3S_{1}$ phase shift, and were reproduced with a relative error $\leq 0.8 \%$. The average of the absolute values of all error values shown in Fig. \ref{fig:error} (a), was calculated to be only $0.23\%$, validating and quantifying the high-accuracy of the calculated potential energy depth and the extracted $^3S_{1}$ phase shift. 
Furthermore, in Fig. \ref{fig:error} (b) we show the relative error between the experimental and theoretical estimates of vibrational energy splitting for both the inner-well states (marked $\blacklozenge$) and the outer-well states.
This serves as a better estimate for the accuracy with which we probe the shape of the potential well.
The difference between adjacent energy levels in both the triplet and the mixed PECs are obtained separately and the relative error is shown for each $n$-value.
States localised in the outer-wells (for $n$=24 to 27) exhibit maximal relative error \SI{2.7}{\percent}, with the inner-well counterparts featuring error rates less than \SI{10}{\percent}. The states localized in the single well of the $n$=22 trilobite potential exhibit error rates less than \SI{5}{\percent}. Further improvements in the molecular Hamiltonian would be necessary to produce numerical results that reproduce the experimental data within the resolution achieved.
\newline

\paragraph{Conclusion:} To summarize, we have measured pure trilobite Rydberg series for five different principal quantum numbers at a resolution of $10^{-4}$.
Permanent electric dipole moments of almost \SI{3000}{debye} were measured and it was shown that this allows an unambiguous assignment of vibrational states in multi-well potentials.
Furthermore, the recently developed Green's function method was used to analyze the most extensive experimental study of the trilobite ULRM to date. This allowed the binding energies and vibrational splittings to be theoretically reproduced with a relative error less than \SI{0.8}{\percent} and \SI{10}{\percent} respectively, providing us with an accurate estimate of the depth and shape of the potential energy curves.
The vibrational ground states measured have large binding energies in comparison to the resolution, and are exceedingly $S$-wave dominant. 
This allowed for the extraction of low energy $^3S_1$-wave scattering phase shift for $^{87}$Rb with unprecedented accuracy in the relevant $k$-range. We argue that the accuracy achieved is largely limited by the predictive power of the Hamiltonian and an enhanced understanding of the molecular interactions is necessary to proceed further. The high resolution data provided is a strong benchmark to test future theoretical models of the ULRM, and can potentially lead to an even deeper insight into low-energy electron-atom scattering.
\newline



\textit{Acknowledgements:} We would like to thank Max Althön for helpful discussions. This project is funded by the German science foundation DFG, project numbers 460443971, 316211972, and INST 248/268-1.
\newline


\textit{Data availability:} The data that support the findings of this study are available from the corresponding author upon reasonable request.

%

\onecolumngrid
\appendix

\section{End Matter}

All bound states localized in the trilobite potential energy curves are predominantly of $S$-wave character and can be used to extract $^3S_1$ phase shifts. However, near the avoided crossing with the butterfly potential curve, the electronic state of the molecule acquires some amount of $^3P_J$ character. 
Hence, large changes in the $^3P_J$ phase shifts change the exact position of the avoided crossing, thereby limiting the number of bound-states observed in the inner well.
Since the outer-most well of the trilobite molecule for $n \geq 24$ is much less affected by large changes in the $P$-wave phase shifts, we first fit the $^3S_1$ phase shift in the range $k \in [0.01,0.018]$. Subsequently, we simultaneously change the $S$-wave and $P$-wave phase shifts to minimize the errors in the inner-well states in the range $k \in [0.018,0.025]$, while avoiding un-physical oscillatory behavior in the $^3S_1$ phase shift.

\begin{figure}[h]
\centering
\includegraphics[width=0.7\linewidth]{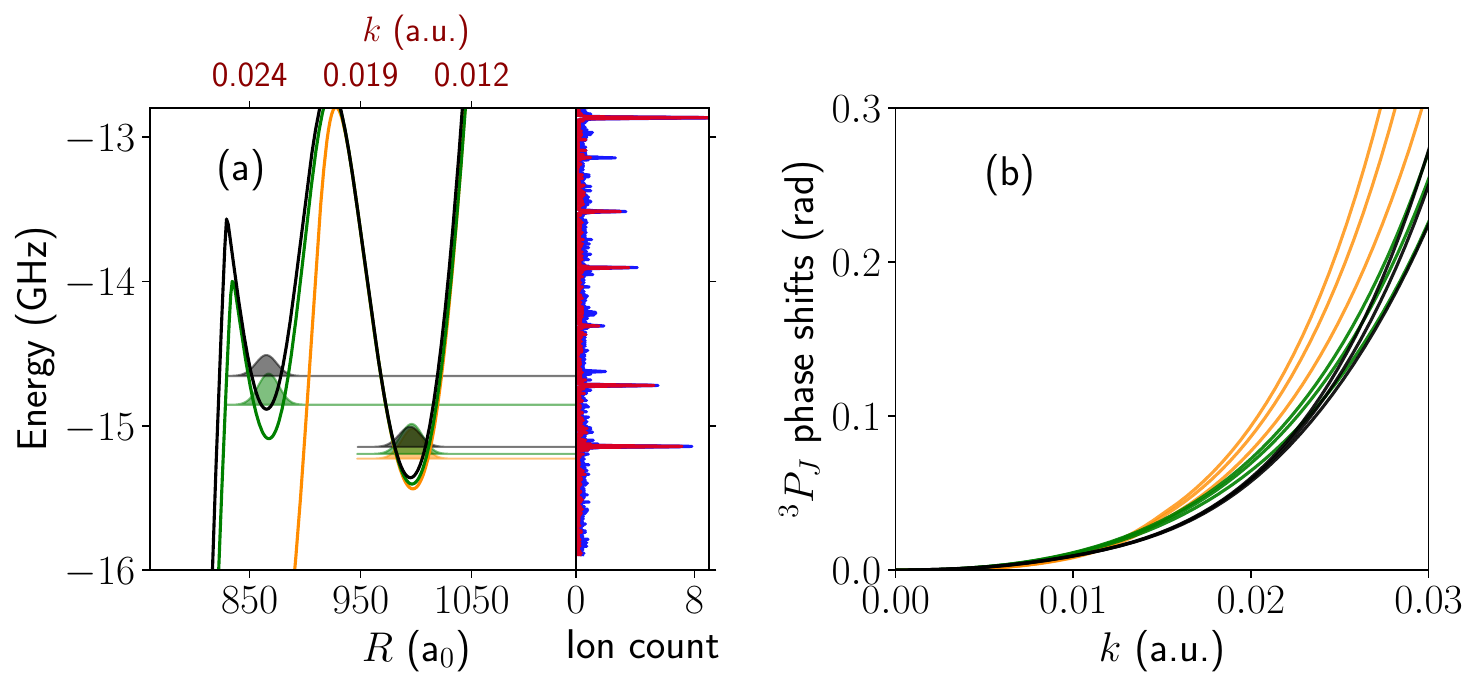}

\caption{The potential energy structure of the $n$=24 trilobite molecule (a) calculated with the newly extracted $^3S_1$ phase shifts with varying $^3P_J$ phase shifts (b). The orange lines are the $^3P_J$ phase shifts taken from ab-initio calculation, the green lines are phase shifts taken from the previous fit, and the black lines are the newly extracted phase shifts.}
\label{fig:p_wave_comparison}
\end{figure}

\begin{figure*}
\centering
\includegraphics[width=1\linewidth]{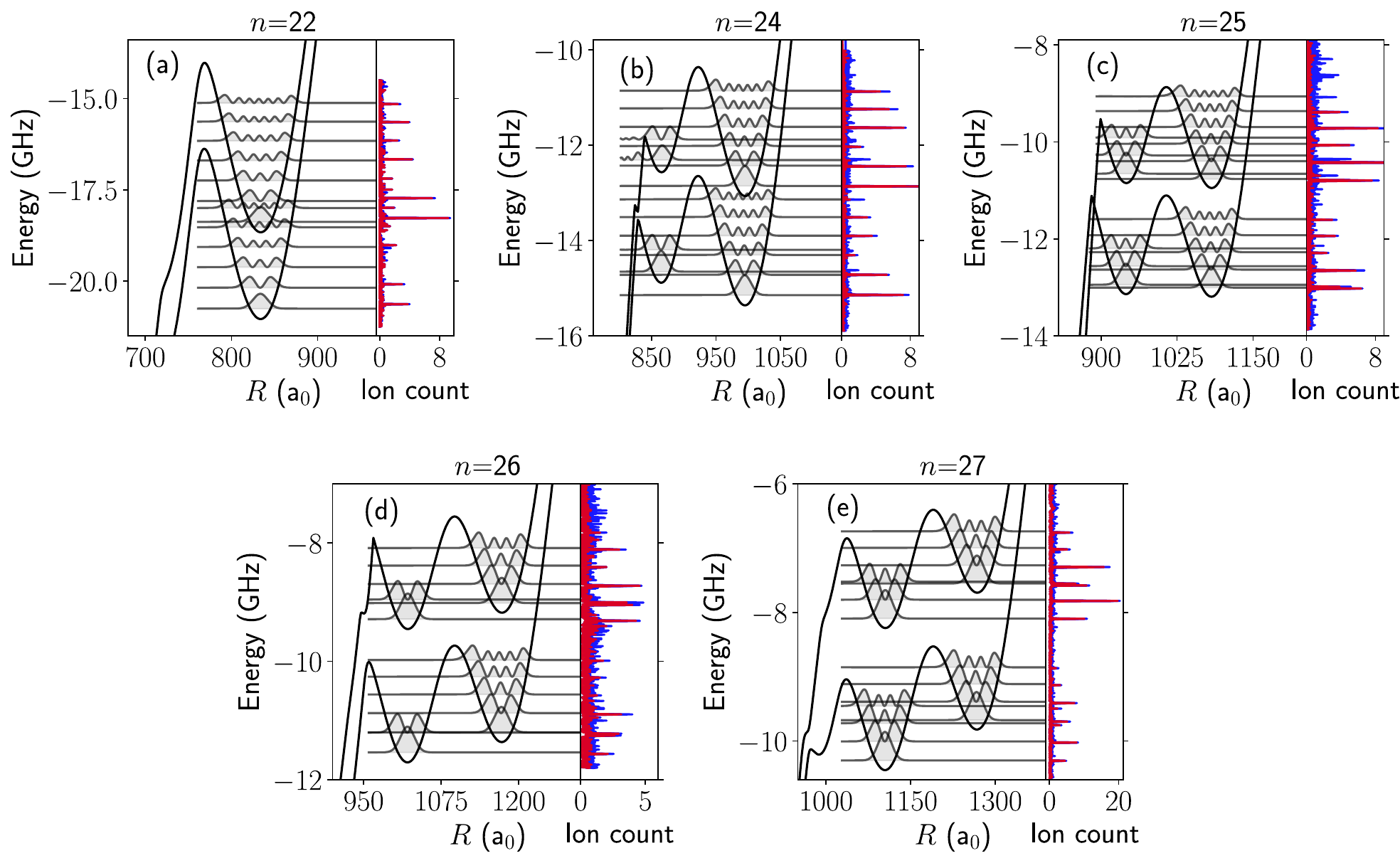}
\caption{The vibrational spectra of the trilobite molecule (black) for all the measured $n$-values, from n=22 to 27 plotted in panels (a) to (e) respectively. The blue (red) lines represent the full- (zero-) momentum ion-counts.} 
\label{fig:n_spectrum}
\end{figure*}

Figure \ref{fig:p_wave_comparison} shows the extracted $^3P_J$ phase shifts (black) and the sensitivity of the potential energy structure to large changes in the $P$-wave phase shifts. For comparison, we show the potential energy curve of the $n$=24 trilobite utilizing different sets of $P$-wave phase shifts. We see that the outer-well bound states are very insensitive to even large changes in $P$-wave phase shifts, as expected from their nearly pure $^3S_1$ character. However, the $P$-wave phase shifts by Ref.~\cite{PhysRevLett.123.073003} (green) do not accurately predict the inner-well bound states, while the phase shifts by Ref.~\cite{Khuskivadze2002} (orange) renders the inner well and the observed inner-well bound-state completely absent. The newly extracted set of phase shifts (black) not only accurately predict the positions of the inner-well and outer-well bound states for $n$=24, it also reproduces the double-well structure for $n$=24-27 and explains all the double peak structures quite well (see Fig.~\ref{fig:n_spectrum}).

Figure \ref{fig:n_spectrum} shows the vibrational spectra of the trilobite molecule for all $n$-values observed. The vibrational states calculated using the newly extracted phase shifts are plotted adjacent to the observed ion count signal, making the assignment of peaks to the inner/outer well straightforward. The zero momentum spectra (red) helps us identify stable bound states and accurately depicts almost all fully bound states in the outer-well. The full momentum spectra on the other hand also contains the zero momentum signal, as well as signals from unstable molecular resonances. Signals that are present in the full momentum spectra but absent in the zero momentum spectra represent these unstable states, either highly excited states in the outer well (see Fig. \ref{fig:n_spectrum} (c))  or states localized in the inner-wells of the $n$=24,25 molecule near the avoided crossings (see Fig. \ref{fig:n_spectrum} (c,d)). Note that some of these unstable states might be significantly affected by non-adiabatic coupling, explaining the relatively larger error in the numerical prediction of their vibrational splittings ($\sim$ \SI{10}{\percent}). Since large changes in the $^3P_J$ phase shifts are necessary to affect the binding energies of the inner-well states, the exact form of the $^3P_J$ phase shifts extracted have limited accuracy. Nevertheless the binding energies of all bound states (in the inner and outer-wells) are reproduced very accurately (error $\sim$ \SI{0.8}{\percent}, see Fig.~\ref{fig:error}) as is visible in Fig.~\ref{fig:n_spectrum}.


\begin{thebibliography}{35}%
\makeatletter
\providecommand \@ifxundefined [1]{%
 \@ifx{#1\undefined}
}%
\providecommand \@ifnum [1]{%
 \ifnum #1\expandafter \@firstoftwo
 \else \expandafter \@secondoftwo
 \fi
}%
\providecommand \@ifx [1]{%
 \ifx #1\expandafter \@firstoftwo
 \else \expandafter \@secondoftwo
 \fi
}%
\providecommand \natexlab [1]{#1}%
\providecommand \enquote  [1]{``#1''}%
\providecommand \bibnamefont  [1]{#1}%
\providecommand \bibfnamefont [1]{#1}%
\providecommand \citenamefont [1]{#1}%
\providecommand \href@noop [0]{\@secondoftwo}%
\providecommand \href [0]{\begingroup \@sanitize@url \@href}%
\providecommand \@href[1]{\@@startlink{#1}\@@href}%
\providecommand \@@href[1]{\endgroup#1\@@endlink}%
\providecommand \@sanitize@url [0]{\catcode `\\12\catcode `\$12\catcode
  `\&12\catcode `\#12\catcode `\^12\catcode `\_12\catcode `\%12\relax}%
\providecommand \@@startlink[1]{}%
\providecommand \@@endlink[0]{}%
\providecommand \url  [0]{\begingroup\@sanitize@url \@url }%
\providecommand \@url [1]{\endgroup\@href {#1}{\urlprefix }}%
\providecommand \urlprefix  [0]{URL }%
\providecommand \Eprint [0]{\href }%
\providecommand \doibase [0]{https://doi.org/}%
\providecommand \selectlanguage [0]{\@gobble}%
\providecommand \bibinfo  [0]{\@secondoftwo}%
\providecommand \bibfield  [0]{\@secondoftwo}%
\providecommand \translation [1]{[#1]}%
\providecommand \BibitemOpen [0]{}%
\providecommand \bibitemStop [0]{}%
\providecommand \bibitemNoStop [0]{.\EOS\space}%
\providecommand \EOS [0]{\spacefactor3000\relax}%
\providecommand \BibitemShut  [1]{\csname bibitem#1\endcsname}%
\let\auto@bib@innerbib\@empty




\bibitem [{\citenamefont {Greene}\ \emph {et~al.}(2000)\citenamefont {Greene},
  \citenamefont {Dickinson},\ and\ \citenamefont {Sadeghpour}}]{Greene2000}%
  \BibitemOpen
  \bibfield  {author} {\bibinfo {author} {\bibfnamefont {C.~H.}\ \bibnamefont
  {Greene}}, \bibinfo {author} {\bibfnamefont {A.~S.}\ \bibnamefont
  {Dickinson}},\ und\ \bibinfo {author} {\bibfnamefont {H.~R.}\ \bibnamefont
  {Sadeghpour}},\ }\bibfield  {title} {\bibinfo {title} {Creation of polar and
  nonpolar ultra-long-range {Rydberg} molecules},\ }\href
  {https://doi.org/10.1103/PhysRevLett.85.2458} {\bibfield  {journal} {\bibinfo
   {journal} {Phys. Rev. Lett.}\ }\textbf {\bibinfo {volume} {85}},\ \bibinfo
  {pages} {2458} (\bibinfo {year} {2000})}\BibitemShut {NoStop}%
\bibitem [{\citenamefont {Rosario}(2015)}]{Rosario_2015}%
  \BibitemOpen
  \bibfield  {author} {\bibinfo {author} {\bibfnamefont {J.~A.}\ \bibnamefont
  {Fernández}},\ \bibinfo {author} {\bibfnamefont {H.~R}\ \bibnamefont
  {Sadeghpour}},\ \bibinfo {author} {\bibfnamefont {P.}\ \bibnamefont
  {Schmelcher}},\ und \bibinfo {author} {\bibfnamefont {R.~G}\ \bibnamefont
  {Férez}},\  } \bibfield  {title} {\bibinfo {title} {Ultralong-Range Rb-KRb Rydberg Molecules: Selected Aspects of Electronic Structure, Orientation and Alignment},\ }\href
  {https://dx.doi.org/10.1088/1742-6596/635/1/012023} {\bibfield  {journal} {\bibinfo
  {journal} {Journal of Physics: Conference Series}\
  }\textbf {\bibinfo {volume} {635}},\ \bibinfo {pages} {012023} (\bibinfo
  {year} {2015})}\BibitemShut {NoStop}%
\bibitem [{\citenamefont {Mayle}(2012)}]{Mayle_2012}%
  \BibitemOpen
  \bibfield  {author} {\bibinfo {author} {\bibfnamefont {M.}\ \bibnamefont
  {Mayle}},\ \bibinfo {author} {\bibfnamefont {S.~T.}\ \bibnamefont
  {Rittenhouse}},\ \bibinfo {author} {\bibfnamefont {P.}\ \bibnamefont
  {Schmelcher}},\ und \bibinfo {author} {\bibfnamefont {H.~R.}\ \bibnamefont
  {Sadeghpour}},\ } \bibfield  {title} {\bibinfo {title} {Electric field control in ultralong-range triatomic polar Rydberg molecules},\ }\href
  {https://link.aps.org/doi/10.1103/PhysRevA.85.052511} {\bibfield  {journal} {\bibinfo
  {journal} {Phys. Rev. A}\
  }\textbf {\bibinfo {volume} {85}},\ \bibinfo {pages} {052511} (\bibinfo
  {year} {2012})}\BibitemShut {NoStop}%
\bibitem [{\citenamefont {Hollerith}(2023)}]{Hollerith_2023}%
  \BibitemOpen
  \bibfield  {author} {\bibinfo {author} {\bibfnamefont {J.}\ \bibnamefont
  {Zeiher}},\ und \bibinfo {author} {\bibfnamefont {S.}\ \bibnamefont
  {Hollerith}},\ } \bibfield  {title} {\bibinfo {title} {Rydberg Macrodimers: Diatomic Molecules on the Micrometer Scale},\ }\href
  {https://doi.org/10.1021/acs.jpca.2c08454} {\bibfield  {journal} {\bibinfo
  {journal} {J. Phys. Chem. A}\
  }\textbf {\bibinfo {volume} {127}},\ \bibinfo {pages} {3925} (\bibinfo
  {year} {2023})}\BibitemShut {NoStop}%
\bibitem [{\citenamefont {Raithel}(2021)}]{Raithel_2021}%
  \BibitemOpen
  \bibfield  {author} {\bibinfo {author} {\bibfnamefont {A.}\ \bibnamefont
  {Duspayev}},\ \bibinfo {author} {\bibfnamefont {X.}\ \bibnamefont
  {Han}},\ \bibinfo {author} {\bibfnamefont {M.~A.}\ \bibnamefont
  {Viray}},\ \bibinfo {author} {\bibfnamefont {L.}\ \bibnamefont
  {Ma}},\  \bibinfo {author} {\bibfnamefont {L.}\ \bibnamefont
  {Zhao}},\ und \bibinfo {author} {\bibfnamefont {G.}\ \bibnamefont
  {Raithel}},\ } \bibfield  {title} {\bibinfo {title} {Long-range Rydberg-atom--ion molecules of Rb and Cs},\ }\href
  {https://link.aps.org/doi/10.1103/PhysRevResearch.3.023114} {\bibfield  {journal} {\bibinfo
  {journal} {Phys. Rev. Res.}\
  }\textbf {\bibinfo {volume} {3}},\ \bibinfo {pages} {023114} (\bibinfo
  {year} {2021})}\BibitemShut {NoStop}%
\bibitem [{\citenamefont {Zuber}(2022)}]{Zuber_2022}%
  \BibitemOpen
  \bibfield  {author} {\bibinfo {author} {\bibfnamefont {N.}\ \bibnamefont
  {Zuber}},\ \bibinfo {author} {\bibfnamefont {V. S. V.}\ \bibnamefont
  {Anasuri}},\ \bibinfo {author} {\bibfnamefont {M.}\ \bibnamefont
  {Berngruber}},\ \bibinfo {author} {\bibfnamefont {Y.~Q}\ \bibnamefont
  {Zou}},\  \bibinfo {author} {\bibfnamefont {R.}\ \bibnamefont
  {L{\"o}w}},\ und \bibinfo {author} {\bibfnamefont {T.}\ \bibnamefont
  {Pfau}},\ }\bibfield  {title} {\bibinfo {title} {Observation of a molecular bond between ions and Rydberg atoms},\ }\href
  {https://doi.org/10.1038/s41586-022-04577-5} {\bibfield  {journal} {\bibinfo
  {journal} {Nature}\
  }\textbf {\bibinfo {volume} {52}},\ \bibinfo {pages} {453} (\bibinfo
  {year} {2022})}\BibitemShut {NoStop}%
\bibitem [{\citenamefont {Dan}(2023)}]{Dan_2023}%
  \BibitemOpen
  \bibfield  {author} {\bibinfo {author} {\bibfnamefont {D.}\ \bibnamefont
  {Bosworth}},\ \bibinfo {author} {\bibfnamefont {F.}\ \bibnamefont
  {Hummel}},\ und \bibinfo {author} {\bibfnamefont {P.}\ \bibnamefont
  {Schmelcher}},\ }\bibfield  {title} {\bibinfo {title} {Charged ultralong-range Rydberg trimers},\ }\href
  {https://link.aps.org/doi/10.1103/PhysRevA.107.022807} {\bibfield  {journal} {\bibinfo
  {journal} {Physics. Rev. A}\
  }\textbf {\bibinfo {volume} {107}},\ \bibinfo {pages} {022807} (\bibinfo
  {year} {2023})}\BibitemShut {NoStop}%
\bibitem [{\citenamefont {Eiles}(2019)}]{Eiles_2019}%
  \BibitemOpen
  \bibfield  {author} {\bibinfo {author} {\bibfnamefont {M.~T.}\ \bibnamefont
  {Eiles}},\ }\bibfield  {title} {\bibinfo {title} {Trilobites, butterflies,
  and other exotic specimens of long-range {Rydberg} molecules},\ }\href
  {https://doi.org/10.1088/1361-6455/ab19ca} {\bibfield  {journal} {\bibinfo
  {journal} {Journal of Physics B: Atomic, Molecular and Optical Physics}\
  }\textbf {\bibinfo {volume} {52}},\ \bibinfo {pages} {113001} (\bibinfo
  {year} {2019})}\BibitemShut {NoStop}%
\bibitem [{\citenamefont {Fey}(2020)}]{Fey_2020_review}%
  \BibitemOpen
  \bibfield  {author} {\bibinfo {author} {\bibfnamefont {C.}\ \bibnamefont
  {Fey}},\ \bibinfo {author} {\bibfnamefont {F.}\ \bibnamefont
  {Hummel}},\ und \bibinfo {author} {\bibfnamefont {P.}\ \bibnamefont
  {Schmelcher}},\ } \bibfield  {title} {\bibinfo {title} {Ultralong-range Rydberg molecules},\ }\href
  {https://doi.org/10.1080/00268976.2019.1679401} {\bibfield  {journal} {\bibinfo
  {journal} {Mol. Phys.}\
  }\textbf {\bibinfo {volume} {118}},\ \bibinfo {pages} {e1679401} (\bibinfo
  {year} {2020})}\BibitemShut {NoStop}%
\bibitem [{\citenamefont {Dunning}(2024)}]{Dunning_2024_review}%
  \BibitemOpen
  \bibfield  {author} {\bibinfo {author} {\bibfnamefont {F.~B.}\ \bibnamefont
  {Dunning}},\ \bibinfo {author} {\bibfnamefont {S.~K.}\ \bibnamefont
  {Kanungo}},\ \bibinfo {author} {\bibfnamefont {S.}\ und \bibnamefont
  {Yoshida}},\ } \bibfield  {title} {\bibinfo {title} {Ultralong-range Rydberg molecules},\ }\href
  {https://dx.doi.org/10.1088/1361-6455/ad7459} {\bibfield  {journal} {\bibinfo
  {journal} {J. Phys. B: At. Mol. Opt. Phys.}\
  }\textbf {\bibinfo {volume} {57}},\ \bibinfo {pages} {212002} (\bibinfo
  {year} {2024})}\BibitemShut {NoStop}%
\bibitem [{\citenamefont {Hamilton}\ \emph {et~al.}(2002)\citenamefont
  {Hamilton}, \citenamefont {Greene},\ und\ \citenamefont {Sadeghpour}}]{Hamilton_2002}%
  \BibitemOpen
  \bibfield  {author} {\bibinfo {author} {\bibfnamefont {E.~L.}~\bibnamefont
  {Hamilton}}, \bibinfo {author} {\bibfnamefont {C.~H.}~\bibnamefont {Greene}},\ und\
  \bibinfo {author} {\bibfnamefont {H.~R.}~\bibnamefont {Sadeghpour}},\ }\bibfield
  {title} {\bibinfo {title} {Shape-resonance-induced long-range molecular Rydberg states},\ }\href
  {https://dx.doi.org/10.1088/0953-4075/35/10/102} {\bibfield  {journal} {\bibinfo
  {journal} {Journal of Physics B: Atomic, Molecular and Optical Physics}\ }\textbf {\bibinfo {volume} {35}},\
  \bibinfo {pages} {L199} (\bibinfo {year} {2002})}\BibitemShut {NoStop}%
 \bibitem [{\citenamefont {Li}\ \emph {et~al.}(2011)\citenamefont
  {Li}, \citenamefont {Pohl}, \citenamefont {Rost}, \citenamefont {Rittenhouse}, \citenamefont {Sadeghpour}, \citenamefont {Nipper}, \citenamefont {Butscher}, \citenamefont {Balewski}, \citenamefont {Bendkowsky}, \citenamefont {Löw},\ und\
  \citenamefont {Pfau}}]{Li2011}%
  \BibitemOpen
  \bibfield  {author} {\bibinfo {author} {\bibfnamefont {W.}~\bibnamefont
  {Li}}, \bibinfo {author} {\bibfnamefont {T.}~\bibnamefont {Pohl}},
  \bibinfo {author} {\bibfnamefont {J.~M.}~\bibnamefont {Rost}}, \bibinfo {author} {\bibfnamefont {Seth~T.}~\bibnamefont {Rittenhouse}},
  \bibinfo {author} {\bibfnamefont {H.~R.}~\bibnamefont {Sadeghpour}}, \bibinfo {author} {\bibfnamefont {J.}~\bibnamefont {Nipper}},
  \bibinfo {author} {\bibfnamefont {B.}~\bibnamefont {Butscher}}, \bibinfo {author} {\bibfnamefont {J.~B.}~\bibnamefont {Balewski}},
  \bibinfo {author} {\bibfnamefont {V.}~\bibnamefont {Bendkowsky}}, \bibinfo {author} {\bibfnamefont {R.}~\bibnamefont {Löw}},\ und\
  \bibinfo {author} {\bibfnamefont {T.}~\bibnamefont {Pfau}},\ }\bibfield
  {title} {\bibinfo {title} {A Homonuclear Molecule with a Permanent Electric Dipole Moment},\ }\href
  {https://www.science.org/doi/abs/10.1126/science.1211255} {\bibfield  {journal} {\bibinfo
  {journal} {Science}\ }\textbf {\bibinfo {volume} {334}},\
  \bibinfo {pages} {1110--1114} (\bibinfo {year} {2011})}\BibitemShut {NoStop}%
\bibitem [{\citenamefont {Chibisov}(2002)}]{Chibisov_2002}%
  \BibitemOpen
  \bibfield  {author} {\bibinfo {author} {\bibfnamefont {M.~I.}\ \bibnamefont
  {Chibisov}},\ \bibinfo {author} {\bibfnamefont {A.~A.}\ \bibnamefont
  {Khuskivadze}},\ und \bibinfo {author} {\bibfnamefont {I.~I.}\ \bibnamefont
  {Fabrikant}},\ }\bibfield  {title} {\bibinfo {title} {Energies and dipole moments of long-range molecular Rydberg states},\ }\href
  {https://dx.doi.org/10.1088/0953-4075/35/10/101} {\bibfield  {journal} {\bibinfo
  {journal} {J. Phys. B: At. Mol. Opt. Phys.}\
  }\textbf {\bibinfo {volume} {35}},\ \bibinfo {pages} {L193} (\bibinfo
  {year} {2002})}\BibitemShut {NoStop}%
\bibitem [{\citenamefont {Booth}\ \emph {et~al.}(2015)\citenamefont
  {Booth}, \citenamefont {Rittenhouse}, \citenamefont {Yang}, \citenamefont {Sadeghpour},\ und\
  \citenamefont {Shaffer}}]{Booth2015}%
  \BibitemOpen
  \bibfield  {author} {\bibinfo {author} {\bibfnamefont {D.}~\bibnamefont
  {Booth}}, \bibinfo {author} {\bibfnamefont {S.~T.}~\bibnamefont {Rittenhouse}},
  \bibinfo {author} {\bibfnamefont {J.}~\bibnamefont {Yang}}, \bibinfo {author} {\bibfnamefont {H.~R.}~\bibnamefont {Sadeghpour}},\ und\
  \bibinfo {author} {\bibfnamefont {J.~P.}~\bibnamefont {Shaffer}},\ }\bibfield
  {title} {\bibinfo {title} {Production of trilobite Rydberg molecule dimers with kilo-Debye permanent electric dipole moments},\ }\href
  {https://doi.org/10.1126/science.1260722} {\bibfield  {journal} {\bibinfo
  {journal} {Science}\ }\textbf {\bibinfo {volume} {348}},\
  \bibinfo {pages} {99--102} (\bibinfo {year} {2015})}\BibitemShut {NoStop}%
\bibitem [{\citenamefont {Althön}\ \emph {et~al.}(2023)\citenamefont
  {Althön}, \citenamefont {Exner}, \citenamefont {Blättner},\ und\
  \citenamefont {Ott}}]{Althoen2023}%
  \BibitemOpen
  \bibfield  {author} {\bibinfo {author} {\bibfnamefont {M.}~\bibnamefont
  {Althön}}, \bibinfo {author} {\bibfnamefont {M.}~\bibnamefont {Exner}},
  \bibinfo {author} {\bibfnamefont {R.}~\bibnamefont {Blättner}},\ und\
  \bibinfo {author} {\bibfnamefont {H.}~\bibnamefont {Ott}},\ }\bibfield
  {title} {\bibinfo {title} {Exploring the vibrational series of pure trilobite Rydberg molecules},\ }\href
  {https://doi.org/10.1038/s41467-023-43818-7} {\bibfield  {journal} {\bibinfo
  {journal} {Nature Communications}\ }\textbf {\bibinfo {volume} {14}},\
  \bibinfo {pages} {8108} (\bibinfo {year} {2023})}\BibitemShut {NoStop}%
 \bibitem [{\citenamefont {Engel}\ \emph {et~al.}(2023)\citenamefont
  {Engel}, \citenamefont {Tiwari}, \citenamefont {Pfau}, \citenamefont {Wüster},\ und\
  \citenamefont {Meinert}}]{engel2023situ}%
  \BibitemOpen
  \bibfield  {author} {\bibinfo {author} {\bibfnamefont {F.}~\bibnamefont
  {Engel}}, \bibinfo {author} {\bibfnamefont {S.~K.}~\bibnamefont {Tiwari}},
  \bibinfo {author} {\bibfnamefont {T.}~\bibnamefont {Pfau}}, \bibinfo {author} {\bibfnamefont {S.}~\bibnamefont {Wüster}},\ und\
  \bibinfo {author} {\bibfnamefont {F.}~\bibnamefont {Meinert}},\ }\bibfield
  {title} {\bibinfo {title} {In situ observation of chemistry in Rydberg molecules within a coherent solvent},\ }\href
  {https://arxiv.org/abs/2308.13762} {\bibfield  {journal} {\bibinfo
  {journal} {arXiv preprint arXiv:2308.13762}\ }(\bibinfo {year} {2023})}\BibitemShut {NoStop}%
\bibitem [{\citenamefont {Hummel}\ \emph
  {et~al.}(2021{\natexlab{a}})\citenamefont {Hummel}, \citenamefont {Keiler},\
  and\ \citenamefont {Schmelcher}}]{trilobite_wave_packet}%
  \BibitemOpen
  \bibfield  {author} {\bibinfo {author} {\bibfnamefont {F.}~\bibnamefont
  {Hummel}}, \bibinfo {author} {\bibfnamefont {K.}~\bibnamefont {Keiler}},\
  und\ \bibinfo {author} {\bibfnamefont {P.}~\bibnamefont {Schmelcher}},\
  }\bibfield  {title} {\bibinfo {title} {Electric-field-induced wave-packet
  dynamics and geometrical rearrangement of trilobite {Rydberg} molecules},\
  }\href {https://doi.org/10.1103/PhysRevA.103.022827} {\bibfield  {journal}
  {\bibinfo  {journal} {Phys. Rev. A}\ }\textbf {\bibinfo {volume} {103}},\
  \bibinfo {pages} {022827} (\bibinfo {year} {2021}{\natexlab{a}})}\BibitemShut
  {NoStop}%
 \bibitem [{\citenamefont {Srikumar}\ \emph {et~al.}(2024)\citenamefont
  {Srikumar}, \citenamefont {Rittenhouse},\ und\ \citenamefont {Schmelcher}}]{srikumar2024}%
  \BibitemOpen
  \bibfield  {author} {\bibinfo {author} {\bibfnamefont {R.}~\bibnamefont
  {Srikumar}}, \bibinfo {author} {\bibfnamefont {S.~T.}~\bibnamefont {Rittenhouse}},\ und\
  \bibinfo {author} {\bibfnamefont {P.}~\bibnamefont {Schmelcher}},\ }\bibfield
  {title} {\bibinfo {title} {Internal diffraction dynamics of trilobite molecules},\ }\href
  {https://arxiv.org/abs/2408.02134} {\bibinfo {howpublished} {arXiv preprint arXiv:2408.02134}} (\bibinfo {year} {2024})\BibitemShut {NoStop}%
\bibitem [{\citenamefont {Kleinbach}\ \emph {et~al.}(2017)\citenamefont
  {Kleinbach}, \citenamefont {Meinert}, \citenamefont {Engel}, \citenamefont {Kwon}, \citenamefont {L\"ow}, \citenamefont {Pfau},\ und\
  \citenamefont {Raithel}}]{PhysRevLett.118.223001}%
  \BibitemOpen
  \bibfield  {author} {\bibinfo {author} {\bibfnamefont {K.~S.}~\bibnamefont
  {Kleinbach}}, \bibinfo {author} {\bibfnamefont {F.}~\bibnamefont {Meinert}},
  \bibinfo {author} {\bibfnamefont {F.}~\bibnamefont {Engel}}, \bibinfo {author} {\bibfnamefont {W.~J.}~\bibnamefont {Kwon}},
  \bibinfo {author} {\bibfnamefont {R.}~\bibnamefont {L\"ow}}, \bibinfo {author} {\bibfnamefont {T.}~\bibnamefont {Pfau}},\ und\
  \bibinfo {author} {\bibfnamefont {G.}~\bibnamefont {Raithel}},\ }\bibfield
  {title} {\bibinfo {title} {Photoassociation of Trilobite Rydberg Molecules via Resonant Spin-Orbit Coupling},\ }\href
  {https://link.aps.org/doi/10.1103/PhysRevLett.118.223001} {\bibfield  {journal} {\bibinfo
  {journal} {Phys. Rev. Lett.}\ }\textbf {\bibinfo {volume} {118}},\
  \bibinfo {pages} {223001} (\bibinfo {year} {2017})}\BibitemShut {NoStop}%
\bibitem [{\citenamefont {Bendkowsky}\ \emph {et~al.}(2009)\citenamefont
  {Bendkowsky}, \citenamefont {Butscher}, \citenamefont {Nipper}, \citenamefont {Shaffer}, \citenamefont {Löw},\ und\
  \citenamefont {Pfau}}]{Bendkowsky2009}%
  \BibitemOpen
  \bibfield  {author} {\bibinfo {author} {\bibfnamefont {V.}~\bibnamefont
  {Bendkowsky}}, \bibinfo {author} {\bibfnamefont {B.}~\bibnamefont {Butscher}},
  \bibinfo {author} {\bibfnamefont {J.}~\bibnamefont {Nipper}}, \bibinfo {author} {\bibfnamefont {J.~P.}~\bibnamefont {Shaffer}},
  \bibinfo {author} {\bibfnamefont {R.}~\bibnamefont {Löw}},\ und\
  \bibinfo {author} {\bibfnamefont {T.}~\bibnamefont {Pfau}},\ }\bibfield
  {title} {\bibinfo {title} {Observation of ultralong-range Rydberg molecules},\ }\href
  {https://doi.org/10.1038/nature07945} {\bibfield  {journal} {\bibinfo
  {journal} {Nature}\ }\textbf {\bibinfo {volume} {458}},\
  \bibinfo {pages} {1005--1008} (\bibinfo {year} {2009})}\BibitemShut {NoStop}%
\bibitem [{\citenamefont {Engel}\ \emph {et~al.}(2019)\citenamefont
  {Engel}, \citenamefont {Dieterle}, \citenamefont {Hummel}, \citenamefont {Fey}, \citenamefont {Schmelcher}, \citenamefont {L\"ow}, \citenamefont {Pfau},\ und\ \citenamefont {Meinert}}]{PhysRevLett.123.073003}%
  \BibitemOpen
  \bibfield  {author} {\bibinfo {author} {\bibfnamefont {F.}~\bibnamefont
  {Engel}}, \bibinfo {author} {\bibfnamefont {T.}~\bibnamefont {Dieterle}},
  \bibinfo {author} {\bibfnamefont {F.}~\bibnamefont {Hummel}}, \bibinfo {author} {\bibfnamefont {C.}~\bibnamefont {Fey}},
  \bibinfo {author} {\bibfnamefont {P.}~\bibnamefont {Schmelcher}}, \bibinfo {author} {\bibfnamefont {R.}~\bibnamefont {L\"ow}},
  \bibinfo {author} {\bibfnamefont {T.}~\bibnamefont {Pfau}},\ und\
  \bibinfo {author} {\bibfnamefont {F.}~\bibnamefont {Meinert}},\ }\bibfield
  {title} {\bibinfo {title} {Precision Spectroscopy of Negative-Ion Resonances in Ultralong-Range Rydberg Molecules},\ }\href
  {https://link.aps.org/doi/10.1103/PhysRevLett.123.073003} {\bibfield  {journal} {\bibinfo
  {journal} {Phys. Rev. Lett.}\ }\textbf {\bibinfo {volume} {123}},\
  \bibinfo {pages} {073003} (\bibinfo {year} {2019})}\BibitemShut {NoStop}%
\bibitem [{\citenamefont {Niederpr\"um}\ \emph {et~al.}(2016)\citenamefont
  {Niederpr\"um}, \citenamefont {Thomas}, \citenamefont {Eichert},\ und\ \citenamefont {Ott}}]{Spinflip}%
  \BibitemOpen
  \bibfield  {author} {\bibinfo {author} {\bibfnamefont {T.}~\bibnamefont
  {Niederpr\"um}}, \bibinfo {author} {\bibfnamefont {O.}~\bibnamefont {Thomas}},
  \bibinfo {author} {\bibfnamefont {T.}~\bibnamefont {Eichert}},\ und\
  \bibinfo {author} {\bibfnamefont {H.}~\bibnamefont {Ott}},\ }\bibfield
  {title} {\bibinfo {title} {Rydberg Molecule-Induced Remote Spin Flips},\ }\href
  {https://link.aps.org/doi/10.1103/PhysRevLett.117.123002} {\bibfield  {journal} {\bibinfo
  {journal} {Phys. Rev. Lett.}\ }\textbf {\bibinfo {volume} {117}},\
  \bibinfo {pages} {123002} (\bibinfo {year} {2016})}\BibitemShut {NoStop}%
\bibitem [{\citenamefont {Niederpr{\"u}m}\ \emph {et~al.}(2016)\citenamefont
  {Niederpr{\"u}m}, \citenamefont {Thomas}, \citenamefont {Eichert},
  \citenamefont {Lippe}, \citenamefont {P{\'e}rez-R{\'\i}os}, \citenamefont
  {Greene},\ and\ \citenamefont {Ott}}]{butterfly}%
  \BibitemOpen
  \bibfield  {author} {\bibinfo {author} {\bibfnamefont {T.}~\bibnamefont
  {Niederpr{\"u}m}}, \bibinfo {author} {\bibfnamefont {O.}~\bibnamefont
  {Thomas}}, \bibinfo {author} {\bibfnamefont {T.}~\bibnamefont {Eichert}},
  \bibinfo {author} {\bibfnamefont {C.}~\bibnamefont {Lippe}}, \bibinfo
  {author} {\bibfnamefont {J.}~\bibnamefont {P{\'e}rez-R{\'\i}os}}, \bibinfo
  {author} {\bibfnamefont {C.~H.}\ \bibnamefont {Greene}},\ und\ \bibinfo
  {author} {\bibfnamefont {H.}~\bibnamefont {Ott}},\ }\bibfield  {title}
  {\bibinfo {title} {Observation of pendular butterfly {Rydberg} molecules},\
  }\href {https://doi.org/10.1038/ncomms12820} {\bibfield  {journal} {\bibinfo
  {journal} {Nature Communications}\ }\textbf {\bibinfo {volume} {7}},\
  \bibinfo {pages} {12820} (\bibinfo {year} {2016})}\BibitemShut {NoStop}%
\bibitem [{\citenamefont {Deiss}\ \emph {et~al.}(2020)\citenamefont
  {Saßmannshausen}, \citenamefont {Merkt},\ und\ \citenamefont {Deiglmayr}}]{Deiss_2020}%
  \BibitemOpen
  \bibfield  {author} {\bibinfo {author} {\bibfnamefont {M.}~\bibnamefont
  {Dei\ss{}}}, \bibinfo {author} {\bibfnamefont {S.}~\bibnamefont {Haze}},
  \bibinfo {author} {\bibfnamefont {J.}~\bibnamefont {Wolf}},\ \bibinfo {author} {\bibfnamefont {L.}~\bibnamefont {Wang}},
  \bibinfo {author} {\bibfnamefont {F.}~\bibnamefont {Meinert}},\ \bibinfo {author} {\bibfnamefont {C.}~\bibnamefont {Fey}},
  \bibinfo {author} {\bibfnamefont {F.}~\bibnamefont {Hummel}},\   \bibinfo {author} {\bibfnamefont {P.}~\bibnamefont {Schmelcher}},\ und 
  \bibinfo {author} {\bibfnamefont {J.~H.}~\bibnamefont {Denschlag}},\ }\bibfield
  {title} {\bibinfo {title} {Observation of spin-orbit-dependent electron scattering using long-range Rydberg molecules},\ }\href
  {https://link.aps.org/doi/10.1103/PhysRevResearch.2.013047} {\bibfield  {journal} {\bibinfo
  {journal} {Phys. Rev. Res.}\ }\textbf {\bibinfo {volume} {2}},\
  \bibinfo {pages} {013047} (\bibinfo {year} {2020})}\BibitemShut {NoStop}%
\bibitem [{\citenamefont {Saßmannshausen}\ \emph {et~al.}(2015)\citenamefont
  {Saßmannshausen}, \citenamefont {Merkt},\ und\ \citenamefont {Deiglmayr}}]{PhysRevLett.114.133201}%
  \BibitemOpen
  \bibfield  {author} {\bibinfo {author} {\bibfnamefont {H.}~\bibnamefont
  {Saßmannshausen}}, \bibinfo {author} {\bibfnamefont {F.}~\bibnamefont {Merkt}},\ und 
  \bibinfo {author} {\bibfnamefont {J.}~\bibnamefont {Deiglmayr}},\ }\bibfield
  {title} {\bibinfo {title} {Experimental Characterization of Singlet Scattering Channels in Long-Range Rydberg Molecules},\ }\href
  {https://link.aps.org/doi/10.1103/PhysRevLett.114.133201} {\bibfield  {journal} {\bibinfo
  {journal} {Phys. Rev. Lett.}\ }\textbf {\bibinfo {volume} {114}},\
  \bibinfo {pages} {133201} (\bibinfo {year} {2015})}\BibitemShut {NoStop}%
 \bibitem [{\citenamefont {Peper}\ und\ \citenamefont {Deiglmayr}(2020)}]{PhysRevA.102.062819}%
  \BibitemOpen
  \bibfield  {author} {\bibinfo {author} {\bibfnamefont {M.}~\bibnamefont
  {Peper}}\ und\ \bibinfo {author} {\bibfnamefont {J.}~\bibnamefont {Deiglmayr}},\ }\bibfield
  {title} {\bibinfo {title} {Photodissociation of long-range Rydberg molecules},\ }\href
  {https://link.aps.org/doi/10.1103/PhysRevA.102.062819} {\bibfield  {journal} {\bibinfo
  {journal} {Phys. Rev. A}\ }\textbf {\bibinfo {volume} {102}},\
  \bibinfo {pages} {062819} (\bibinfo {year} {2020})}\BibitemShut {NoStop}%
\bibitem [{\citenamefont {Peper}\ und\ \citenamefont {Deiglmayr}(2021)}]{PhysRevLett.126.013001}%
  \BibitemOpen
  \bibfield  {author} {\bibinfo {author} {\bibfnamefont {M.}~\bibnamefont
  {Peper}}\ und\ \bibinfo {author} {\bibfnamefont {J.}~\bibnamefont {Deiglmayr}},\ }\bibfield
  {title} {\bibinfo {title} {Heteronuclear Long-Range Rydberg Molecules},\ }\href
  {https://link.aps.org/doi/10.1103/PhysRevLett.126.013001} {\bibfield  {journal} {\bibinfo
  {journal} {Phys. Rev. Lett.}\ }\textbf {\bibinfo {volume} {126}},\
  \bibinfo {pages} {013001} (\bibinfo {year} {2021})}\BibitemShut {NoStop}%
\bibitem [{\citenamefont {MacLennan}\ \emph {et~al.}(2019)\citenamefont
  {MacLennan}, \citenamefont {Chen},\ und\ \citenamefont {Raithel}}]{PhysRevA.99.033407}%
  \BibitemOpen
  \bibfield  {author} {\bibinfo {author} {\bibfnamefont {J.~L.}~\bibnamefont
  {MacLennan}}, \bibinfo {author} {\bibfnamefont {Y.-J.}~\bibnamefont {Chen}},\ und\
  \bibinfo {author} {\bibfnamefont {G.}~\bibnamefont {Raithel}},\ }\bibfield
  {title} {\bibinfo {title} {Deeply bound ($24{D}_{J}+5{S}_{1/2}$) $^{87}\mathrm{Rb}$ and $^{85}\mathrm{Rb}$ molecules for eight spin couplings},\ }\href
  {https://link.aps.org/doi/10.1103/PhysRevA.99.033407} {\bibfield  {journal} {\bibinfo
  {journal} {Phys. Rev. A}\ }\textbf {\bibinfo {volume} {99}},\
  \bibinfo {pages} {033407} (\bibinfo {year} {2019})}\BibitemShut {NoStop}%
\bibitem [{\citenamefont {Greene}\ \emph {et~al.}(2023)\citenamefont
  {Greene},\ und\ \citenamefont {Eiles}}]{green_function}%
  \BibitemOpen
  \bibfield  {author} {\bibinfo {author} {\bibfnamefont {C.~H.}~\bibnamefont
  {Greene}},\ und\ \bibinfo {author} {\bibfnamefont {M.~T.}~\bibnamefont {Eiles}},\ }\bibfield
  {title} {\bibinfo {title} {Green's-function treatment of Rydberg molecules with spins},\ }\href
  {https://link.aps.org/doi/10.1103/PhysRevA.108.042805} {\bibfield  {journal} {\bibinfo
  {journal} {Phys. Rev. A}\ }\textbf {\bibinfo {volume} {108}},\
  \bibinfo {pages} {042805} (\bibinfo {year} {2023})}\BibitemShut {NoStop}%
\bibitem [{\citenamefont {Hummel}\ \emph {et~al.}(2023)\citenamefont
  {Hummel}, \citenamefont {Schmelcher},\ und\ \citenamefont {Eiles}}]{VibronicRb}%
  \BibitemOpen
  \bibfield  {author} {\bibinfo {author} {\bibfnamefont {F.}~\bibnamefont
  {Hummel}}, \bibinfo {author} {\bibfnamefont {P.}~\bibnamefont {Schmelcher}},\ und\
  \bibinfo {author} {\bibfnamefont {M.~T.}~\bibnamefont {Eiles}},\ }\bibfield
  {title} {\bibinfo {title} {Vibronic interactions in trilobite and butterfly Rydberg molecules},\ }\href
  {https://link.aps.org/doi/10.1103/PhysRevResearch.5.013114} {\bibfield  {journal} {\bibinfo
  {journal} {Phys. Rev. Res.}\ }\textbf {\bibinfo {volume} {5}},\
  \bibinfo {pages} {013114} (\bibinfo {year} {2023})}\BibitemShut {NoStop}%
\bibitem [{\citenamefont {Srikumar}\ \emph {et~al.}(2023)\citenamefont
  {Srikumar}, \citenamefont {Hummel},\ und\ \citenamefont {Schmelcher}}]{VibronicNa}%
  \BibitemOpen
  \bibfield  {author} {\bibinfo {author} {\bibfnamefont {R.}~\bibnamefont
  {Srikumar}}, \bibinfo {author} {\bibfnamefont {F.}~\bibnamefont {Hummel}},\ und\
  \bibinfo {author} {\bibfnamefont {P.}~\bibnamefont {Schmelcher}},\ }\bibfield
  {title} {\bibinfo {title} {Nonadiabatic interaction effects in the spectra of ultralong-range Rydberg molecules},\ }\href
  {https://link.aps.org/doi/10.1103/PhysRevA.108.012809} {\bibfield  {journal} {\bibinfo
  {journal} {Phys. Rev. A}\ }\textbf {\bibinfo {volume} {108}},\
  \bibinfo {pages} {012809} (\bibinfo {year} {2023})}\BibitemShut {NoStop}%
\bibitem [{\citenamefont {Eiles}\ \emph {et~al.}(2024)\citenamefont
  {Srikumar} }]{KatosMatt}%
  \BibitemOpen
  \bibfield  {author} {\bibinfo {author} {\bibfnamefont {M.~T.}~\bibnamefont
  {Eiles}},\ und \bibinfo {author} {\bibfnamefont {F.}~\bibnamefont {Hummel}},\ }\bibfield
  {title} {\bibinfo {title} {Kato's theorem and ultralong-range Rydberg molecules},\ }\href
  {https://link.aps.org/doi/10.1103/PhysRevA.109.022811} {\bibfield  {journal} {\bibinfo
  {journal} {Phys. Rev. A}\ }\textbf {\bibinfo {volume} {109}},\
  \bibinfo {pages} {022811} (\bibinfo {year} {2024})}\BibitemShut {NoStop}%
\bibitem [{\citenamefont {Eiles}(2017)}]{Eiles_2017}%
  \BibitemOpen
  \bibfield  {author} {\bibinfo {author} {\bibfnamefont {M.~T.}\ \bibnamefont
  {Eiles}},\ und\ \bibinfo{author}{\bibfnamefont {C.~H.}~\bibnamefont
  {Greene}},\ }\bibfield  {title} {\bibinfo {title} {Hamiltonian for the inclusion of spin effects in long-range Rydberg molecules},\ }\href
  {https://doi.org/10.1103/PhysRevA.95.042515} {\bibfield  {journal} {\bibinfo
  {journal} {Phys. Rev. A}\
  }\textbf {\bibinfo {volume} {95}},\ \bibinfo {pages} {042515} (\bibinfo
  {year} {2017})}\BibitemShut {NoStop}%
\bibitem [{\citenamefont {Fermi}(1934)}]{Fermi1934}%
  \BibitemOpen
  \bibfield  {author} {\bibinfo {author} {\bibfnamefont {E.}\ \bibnamefont
  {Fermi}},\ }\bibfield  {title} {\bibinfo {title} {Sopra lo Spostamento per Pressione delle Righe Elevate delle Serie Spettrali},\ }\href
  {https://doi.org/10.1007/BF02959829} {\bibfield  {journal} {\bibinfo
  {journal} {Nuovo Cim.}\
  }\textbf {\bibinfo {volume} {11}},\ \bibinfo {pages} {157} (\bibinfo
  {year} {1934})}\BibitemShut {NoStop}%
\bibitem [{\citenamefont {Omont}(1977)}]{Omont1977}%
  \BibitemOpen
  \bibfield  {author} {\bibinfo {author} {\bibfnamefont {A.}\ \bibnamefont
  {Omont}},\ }\bibfield  {title} {\bibinfo {title} {On the theory of collisions of atoms in rydberg states with neutral particles},\ }\href
  {https://doi.org/10.1051/jphys:0197700380110134300} {\bibfield  {journal} {\bibinfo
  {journal} {J. Phys. France}\
  }\textbf {\bibinfo {volume} {38}},\ \bibinfo {pages} {1343} (\bibinfo
  {year} {1977})}\BibitemShut {NoStop}%
\bibitem [{\citenamefont {Fey}(2015)}]{Fey_2015}%
  \BibitemOpen
  \bibfield  {author} {\bibinfo {author} {\bibfnamefont {C.}\ \bibnamefont
  {Fey}},\ \bibinfo{author}{\bibfnamefont {M.}~\bibnamefont
  {Kurz}},\  \bibinfo{author}{\bibfnamefont {P.}~\bibnamefont
  {Schmelcher}},\   \bibinfo{author}{\bibfnamefont {S.~T.}~\bibnamefont
  {Rittenhouse}},\ und\ \bibinfo{author}{\bibfnamefont {H.~R.}~\bibnamefont
  {Sadeghpour}},\ }\bibfield  {title} {\bibinfo {title} {A comparative analysis of binding in ultralong-range Rydberg molecules},\ }\href
  {https://dx.doi.org/10.1088/1367-2630/17/5/055010} {\bibfield  {journal} {\bibinfo
  {journal} {New Journal of Physics}\
  }\textbf {\bibinfo {volume} {17}},\ \bibinfo {pages} {055010} (\bibinfo
  {year} {2015})}\BibitemShut {NoStop}%
\bibitem [{\citenamefont {Tarana}(2020)}]{PhysRevA.102.062802}%
  \BibitemOpen
  \bibfield  {author} {\bibinfo {author} {\bibfnamefont {M.}~\bibnamefont
  {Tarana}},\ }\bibfield
  {title} {\bibinfo {title} {Long-range Rydberg molecule ${\mathrm{Rb}}_{2}$: Two-electron $R$-matrix calculations at intermediate internuclear distances},\ }\href
  {https://link.aps.org/doi/10.1103/PhysRevA.102.062802} {\bibfield  {journal} {\bibinfo
  {journal} {Phys. Rev. A}\ }\textbf {\bibinfo {volume} {102}},\
  \bibinfo {pages} {062802} (\bibinfo {year} {2020})}\BibitemShut {NoStop}%
 \bibitem [{\citenamefont {Giannakeas}\ \emph {et~al.}(2020)\citenamefont
  {Giannakeas}, \citenamefont {Eiles}, \citenamefont {Robicheaux},\ und\ \citenamefont {Rost}}]{PhysRevA.102.033315}%
  \BibitemOpen
  \bibfield  {author} {\bibinfo {author} {\bibfnamefont {P.}~\bibnamefont
  {Giannakeas}}, \bibinfo {author} {\bibfnamefont {M.~T.}~\bibnamefont {Eiles}},
  \bibinfo {author} {\bibfnamefont {F.}~\bibnamefont {Robicheaux}},\ und\
  \bibinfo {author} {\bibfnamefont {J.~M.}~\bibnamefont {Rost}},\ }\bibfield
  {title} {\bibinfo {title} {Generalized local frame-transformation theory for ultralong-range Rydberg molecules},\ }\href
  {https://link.aps.org/doi/10.1103/PhysRevA.102.033315} {\bibfield  {journal} {\bibinfo
  {journal} {Phys. Rev. A}\ }\textbf {\bibinfo {volume} {102}},\
  \bibinfo {pages} {033315} (\bibinfo {year} {2020})}\BibitemShut {NoStop}%
\bibitem [{\citenamefont {Geppert}\ \emph {et~al.}(2021)\citenamefont
  {Geppert}, \citenamefont {Alth{\"o}n}, \citenamefont {Fichtner},\ und\
  \citenamefont {Ott}}]{State_changing}%
  \BibitemOpen
  \bibfield  {author} {\bibinfo {author} {\bibfnamefont {P.}~\bibnamefont
  {Geppert}}, \bibinfo {author} {\bibfnamefont {M.}~\bibnamefont {Alth{\"o}n}},
  \bibinfo {author} {\bibfnamefont {D.}~\bibnamefont {Fichtner}},\ and\
  \bibinfo {author} {\bibfnamefont {H.}~\bibnamefont {Ott}},\ }\bibfield
  {title} {\bibinfo {title} {Diffusive-like redistribution in state-changing
  collisions between {Rydberg} atoms and ground state atoms},\ }\href
  {https://doi.org/10.1038/s41467-021-24146-0} {\bibfield  {journal} {\bibinfo
  {journal} {Nature Communications}\ }\textbf {\bibinfo {volume} {12}},\
  \bibinfo {pages} {3900} (\bibinfo {year} {2021})}\BibitemShut {NoStop}%
\bibitem [{\citenamefont {Bahrim}(2000)}]{Bahrim2000}%
  \BibitemOpen
  \bibfield  {author} {\bibinfo {author} {\bibfnamefont {C.}\ \bibnamefont
  {Bahrim}},\ \bibinfo {author} {\bibfnamefont {U.}\ \bibnamefont
  {Thumm}},\ }\bibfield  {title} {\bibinfo {title} {Low-lying ${}^{3}{P}^{o}$ and ${}^{3}{S}^{e}$ states of ${\mathrm{Rb}}^{\ensuremath{-}},{\mathrm{Cs}}^{\ensuremath{-}}$, and ${\mathrm{Fr}}^{\ensuremath{-}}$},\ }\href
  {https://link.aps.org/doi/10.1103/PhysRevA.61.022722} {\bibfield  {journal} {\bibinfo
  {journal} {Phys. Rev. A}\
  }\textbf {\bibinfo {volume} {61}},\ \bibinfo {pages} {022722} (\bibinfo
  {year} {2000})}\BibitemShut {NoStop}%
\bibitem [{\citenamefont {Bahrim}(2001)}]{Bahrim2001}%
  \BibitemOpen
  \bibfield  {author} {\bibinfo {author} {\bibfnamefont {C.}\ \bibnamefont
  {Bahrim}},\ \bibinfo {author} {\bibfnamefont {U.}\ \bibnamefont
  {Thumm}},\ \bibinfo {author} {\bibfnamefont {I.~I.}\ \bibnamefont
  {Fabrikant}},\ }\bibfield  {title} {\bibinfo {title} {Negative-ion resonances in cross sections for slow-electron--heavy-alkali-metal-atom scattering},\ }\href
  {https://link.aps.org/doi/10.1103/PhysRevA.63.042710} {\bibfield  {journal} {\bibinfo
  {journal} {Phys. Rev. A}\
  }\textbf {\bibinfo {volume} {63}},\ \bibinfo {pages} {042710} (\bibinfo
  {year} {2001})}\BibitemShut {NoStop}%
\bibitem [{\citenamefont {Khuskivadze}(2002)}]{Khuskivadze2002}%
  \BibitemOpen
  \bibfield  {author} {\bibinfo {author} {\bibfnamefont {A.~A.}\ \bibnamefont
  {Khuskivadze}},\ \bibinfo {author} {\bibfnamefont {M.~I.}\ \bibnamefont
  {Chibisov}},\ \bibinfo {author} {\bibfnamefont {I.~I.}\ \bibnamefont
  {Fabrikant}},\ }\bibfield  {title} {\bibinfo {title} {Adiabatic energy levels and electric dipole moments of Rydberg states of ${\mathrm{Rb}}_{2}$ and ${\mathrm{Cs}}_{2}$ dimers},\ }\href
  {https://link.aps.org/doi/10.1103/PhysRevA.66.042709} {\bibfield  {journal} {\bibinfo
  {journal} {Phys. Rev. A}\
  }\textbf {\bibinfo {volume} {66}},\ \bibinfo {pages} {042709} (\bibinfo
  {year} {2002})}\BibitemShut {NoStop}%
\end{thebibliography}
\end{document}